# Efficient Resource Allocation in Mass Customization based on Service Oriented Architecture

Ali Vatankhah Barenji
Department of Mechanical Engineering Eastern Mediterranean University, Famagusta, North Cyprus
ali.vatankhah@cc.emu.edu.tr

Reza Vatankhah Barenji
Department of Industrial Engineering Hacettepe University, Beytepe Campus Ankara, Turkey
reza.vatankhah@hacettepe.edu.tr

*Abstract*- **Mass customization explains the phenomenon to provide a unique designed products and services to all customer by achieving a high process integration and flexibility. It has been used as a competitive approach by many companies. Adequate resource implementation in mass customization-particularly in terms of resource usage, it is therefore important to meet customer's requirement in terms effective responsiveness and meeting deadlines, at the same time offering high scalability. An architecture for solving the resource allocation issue in mass customized flexible manufacturing system was illustrated, by putting in place a couple of advance reservation systems and scheduling algorithms for effective usage of the product.**

## I. INTRODUCTION

Mass customization uses Flexible Manufacturing System (FMS) to customize products [1]. FMS focuses on integrating the flexibility of individual customization with the low costs of mass production [2]. Without much increase in cost, there is a significant changes in customization and variety. The system has a significant role in implementing mass customization processes. FMS plays a major rule in deploying mass customization paradigm. It is a manufacturing system with multifunctional machines connected by material handling system, which are controlled by a computer system. Scheduling of the FMS is a great problem in mass customization among many other operational problems because it of the decisions to make when system is in operation [3].

Scheduling is a process that helps to allocate limited manufacturing resources over time including sequential and parallel manufacturing activities, which many has researched in the past years[]. An important part of scheduling is the Manufacturing resource allocation, which it's important in optimizing the production of the system, and also reducing the cost of production.

There are different levels for resource allocation, e.g work cell or individual machine, multi-plant companies, supply chain or virtual enterprise, plant or shop-floor[2]. To solve these problems, several methods which includes genetic algorithms, fuzzy algorithms and intelligent agents has been proposed. For examples, a scheduling framework at the shop floor level was proposed by Wang et al [4], they defined shop floor as a collection of several work cells, where the proposed framework is used as part of flexible manufacturing system. A knowledge-based genetic algorithm that integrate expert knowledge and inherent quality of simple genetic algorithm for searching the optima concurrently was proposed by Prakash,Chan and Deshmukh[5], it explains a scheduling problem in flexible manufacturing system. A divide-and-conquer method to calculate the timetable of a given sequence was proposed by Zhu, Li and Shen[6], it explains a two-machine no-wait job shop problem with a minimal make span. Zhang et al[7] illustrated a manufacturing resource allocation approach by using extended genetic algorithm, the method serves as a bridge to a gap of multi-objective decision-making optimization for the deployment of a supply chain. A fuzzy multi-criteria decision-making approach was presented by Kahraman, Beskese and Kaya, this helps for the selection among enterprise resource planning outsourcing alternatives based on the analytic hierarchy process under fuzziness[8]. An integrated constraint programming model that deals with important issues in flexible manufacturing system such as machine loading, tool allocation, scheduling and part routing was illustrated by Zeballos, Quiroga and Henning[9]. Summarily, all research efforts has contributed greatly to the literature, worthy to note is manufacturing on resource allocation for MC, this has enabled further development and more research on manufacturing resource allocation[10-14]. efforts by authors has highlighted problems of resource allocation such as statistical-based resource allocation, static resource allocation, random resource allocation, majorly for tool selection and sequencing in the environments of supply chain, cloud manufacturing and distribution. In other hand, the stated factors above necessitate an effective selection of resources, in order to meet requirements by individual users, requirements such as deadlines and also providing Quality-Of-Service (QoS)[12]. Furthermore, to avoid collapses, improve scalability and to handle higher workloads there is a need for effective resource allocation, 1.e resource scheduling and which resource to use and when to use the resource. This research present an architecture which is service oriented that supports resource allocation in mass customization to customize a product and also to achieve consumer's requirements. Priori reservation algorithms and runtime scheduling algorithms to achieve optimal workflow achieve Scheduling and resource management.

## II. ARCHITECTURE DESCRIPTION

We illustrate the important features for mass customization in implementing our resource allocation, which is stated in Fig 1. At the company's sale platform, consumers tender their customized products order (CP) which will be later forwarded to the scheduler. No task-to-specific resources for the customized products order tendered to the sales portal. Information such as deadline, customer's requirements are contained inside the metadata associated with the CP.

The received CP meads will be examined by the scheduler, and the information gathered will be used with the present state of the material handling systems, machines, buffers, to ensure effective use of the available resources by adding task-to-resource mappings and also some scheduling info. The Storage and Process information services(SI & IS) provides the current state of the stations(machines) and buffers to scheduler. The scheduler present the executable CP metadata to any of the Workflow Execution Engines (WEE) present in the architecture at the planned starting time. Any of the workflow execution engines has an autonomy and can execute workflow by sending each task to its originally selection machines.

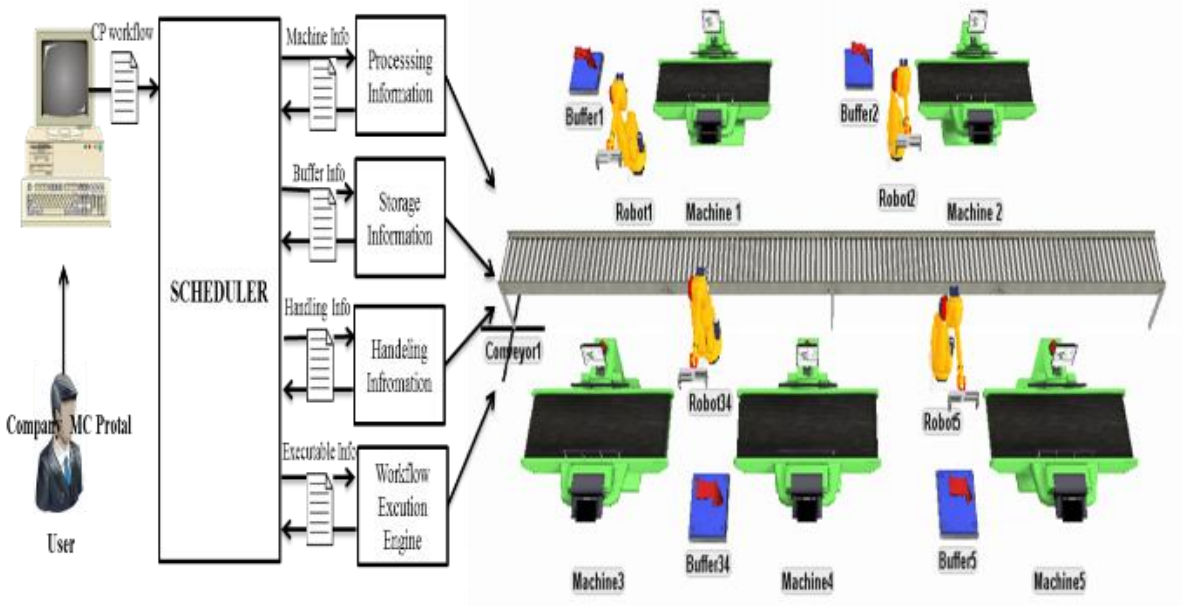

Figure 1 The resource allocation architecture for mass customization

## III. WORKFLOW SCHEDULER: DESIGN DETAILS

The working mechanism of the scheduler is presented in Fig. 2. Directed Acyclic Graph (DAG) is used to represent a customer products workflow, the graph nodes represent atomic tasks and the control dependencies between tasks is represented by the edges. Workflow control structures allows the model to be extended, such control structures includes parallel execution and conditionals. An atomic task functions for performing a computational duty on an input data set and the output data set comprises of the result.

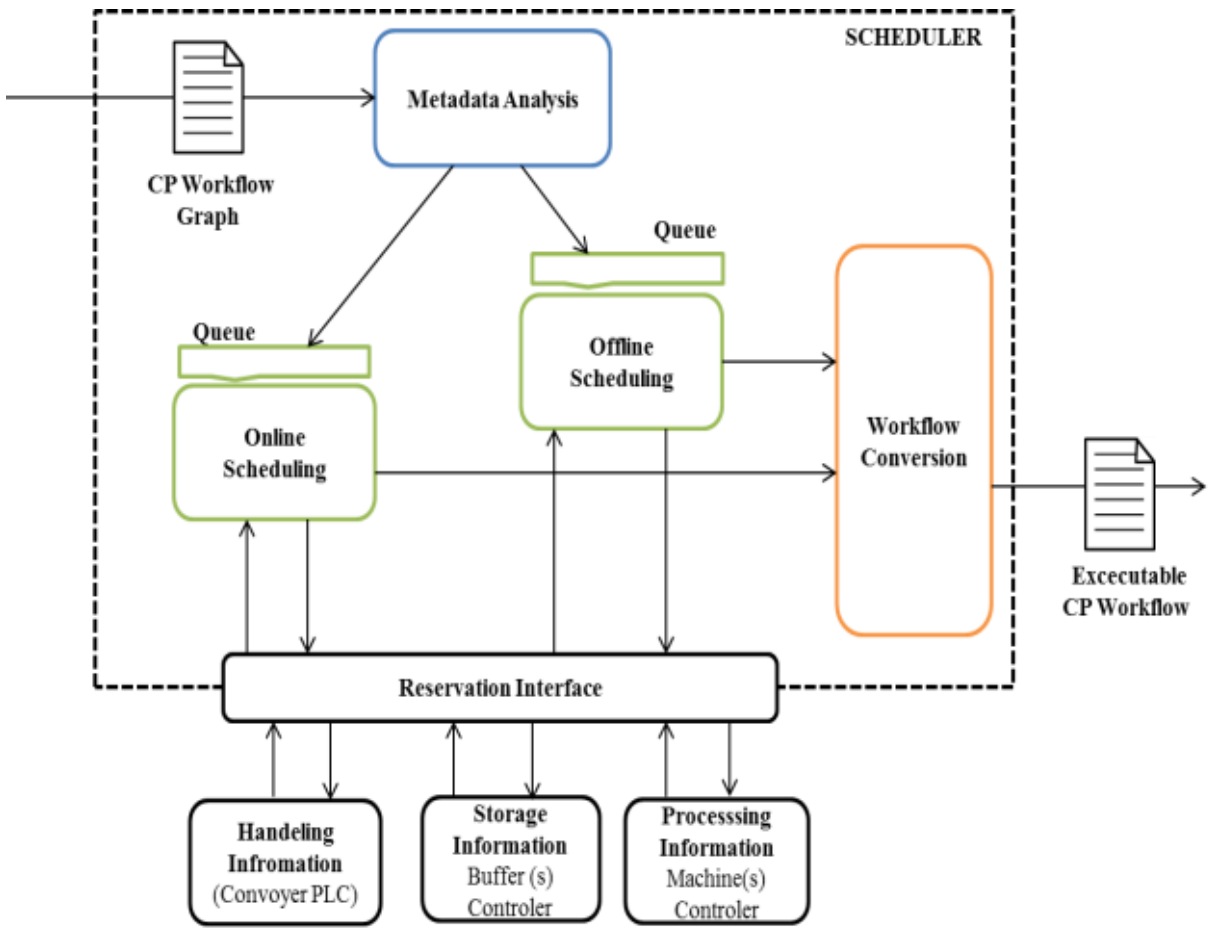

Figure 2 the inner workings mechanism of the Scheduler

The analyzer will examine the CP workflow's metadata, and present on queue the CP workflow to one of the scheduling algorithms. The scheduler has two scheduling algorithms present, which each performing different scheduling method. Online scheduling is used by the first for processing CP workflows in real-time as they are submitted. Manufacturing process however usually have a large number of recurring and predictable workflows. The use of resource can be greatly improved by considering the advantage of this advance knowledge, by the use of offline algorithms to schedule flows. Both approach work together, workflows are scheduled as they arrive in online algorithms while scheduling of workflows is done periodically in offline algorithms.

To integrate the combination of scheduling methods, a system of advance reservations on each resource (machines, buffers and conveyors) is employed, which ensures that resources are available within a given period of time. This method ensures both scheduling approaches work together without greatly affecting the scheduling decision made by each other, as decision made by one scheduling algorithms will be known by the other. In the timeslot tables has buffers and stations on a per-resource basis by the Storage information and Processing information respectively, the conveyer reservations on a per-link basis is handled by the Handling information.

Worthy to note, a solution to the outlined problem of system oversubscription is provided advance reservations, it helps to manage the reservations on the individual resource in a way to solve the problem. We have opted to implement this by granting transfers fully exclusive use of manufacturing system. We decided to use this by granting transfers total exclusive use of manufacturing system. In stations, conveyers are buffers, reservations are requested when a CP workflow is scheduled, this help if a conflicting reservation is rejected and there is a need for workflow to be scheduled for another time or other resources. This approach helps to tackle congestion problem on the resources and suppresses the need for advantage technology such as RFID technology, higher scheduling systems and also distributed manufacturing control. This architecture's requirements and dependencies can be greatly reduced by advance reservations, which tackles the problem of reserving resources for the duration of the whole task, resources are reserved only when needed. This helps in co-scheduling of resources, which improves the overall efficiency of the system. Nevertheless, it complicates the reservation method as inter-and-intra-task dependencies should be taken into consideration. To complete the process, after appropriate resources is selected by the scheduling algorithms and the timing for the CP workflows is determined, an executable CP workflows is produced by decisions of the scheduling algorithms and presented to a workflow engine for process at the required starting time.

## IV. SCHEDULING ALGORITHMS

A non-deterministic polynomial-hard are mapping and scheduling problems [8]. To determine an optimal solution in a given time frame, heuristics approach is required. Suitable resources, reservations for all resources such as station, buffer and conveyor involved, and processes starting time for each CP workflow are to be determined by such algorithms. Some constraints determine the reservation for a workflow. The stations, buffers and each conveyor used for raw and product handling requires some amount of reservations. CP workflow requirements such as operation specification and machine capacity usually determines the length of a single reservation on a resource, reservation length can also be affected by streaming parts on conveyor. As stated earlier, unwanted long reservations should be avoided as we implement a co-scheduling reservation scheme. When the reservation on an input conveyor end, the reservation on the machine should at about same time. Sub-processes can only start after the main processes are completed and reservations on a single resource cannot overlap each other, this was ensured by additional reservation dependencies. Reservations of task on pallet J(i.e input), a buffer i(i.e output) and a manufacturing machine(i.e Machine K) is shown in Fig 3. Operation begins from input pallet and streams and end with the machining process in the output buffer The machine reservations will only start after the input pallet reservation end, this ensure co-scheduling. At the completion of machining reservation, the buffering reservation will start. Only after the last reservation of task $t_i$ will reservation for task $t_j$ allowed as it is the process for the CP workflow.

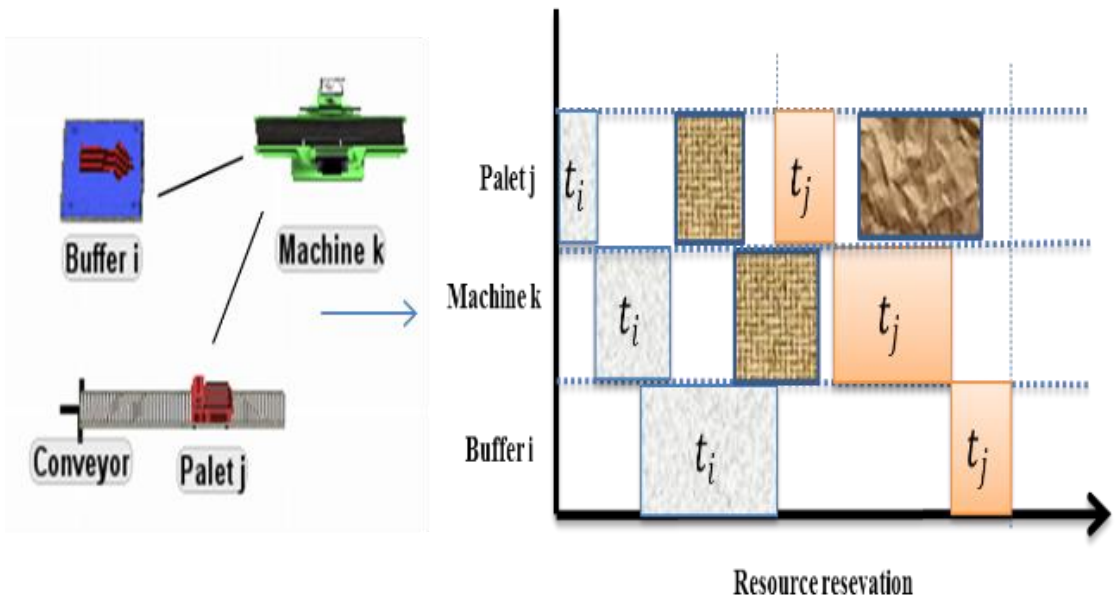

Figure 3 The reservations of task

## V. OFFLINE SCHEDULING ALGORITHMS

We base our newly developed heuristics on the list scheduling technique [5]. Assigning priorities to ordered list of processes is the main aim of list scheduling technique, and also executing the next two steps till a valid schedule is achieved. The "root nodes" of every CP workflow is added to a list and this initialize the algorithm. Priority is designated to each root node in this list and the assignment is giving to the highest ranking feature. This feature is giving to a set of resources (conveyers, stations, buffers,) and valid reservations are giving to the selected resources. There will be action taken to handle a case of no acceptable resource combination found as a result of possible constraint violations. We consider the sub-features of the selected parent feature for insertion in the list finally and the process start all over again until no more features in the list. Different approach are possible for giving these priorities, finding reservations and also constraint violations handling, and by giving tasks to resources. These approach normally have an important impact in the overall efficiency of the algorithms. The assignment of task priority strategy illustrated in this paper are "Closest Deadline (CD)", it assigns higher work to tasks with workflow almost violating a deadline, while Relatively Closest Deadline (RCD)" assign tasks to a closer to violating-a-deadline workflow. Priorities tasks with the lowest free resources by also taken into consideration the load on the resource which is known as |Least Utilization" which includes stations, buffers and conveyors, while the opposite is known as "Most Utilization"

The strategy Detached Selection(DS) for assigning task uses a different resource allocation method, by choosing the stations and machines for the task, and also after selecting the machine, it allocate task to the most available conveyors and buffer. Another selection strategy is the Integrated Selection (IS) that uses a unified resource allocation method, this approach is achieved by assigning a performance check to every resources used for machining and buffering, and the resource combination with the highest performance will be selected for the task. Task requirements, the properties of the resource, optimization objectives and the present reservations on the resources are the criteria for selecting the fitness. Some constraints has been employed for the proposed algorithms. There will be a complete rejection of workflow where there is no schedule that is feasible. When there is a proper scheduling of workflow, there will be less failure of other workflows in meeting their constraints, and important resources are used. If one of more flows are rejection, it doesn't mean all the flows in the batch are rejected.

## VI. ONLINE SCHEDULING ALGORITHMS

List scheduling was presented by the two online heuristics, no priority level is applied to both algorithms. Depending on the present state of the system, scheduling is done the moment each workflow is submitted. There is an evaluation of each combination of conveyor, buffering resources and machining by the Detached Selection (DS) algorithms, the evaluation is done based on the current load and their properties. The method presented in Section 3.1 is similar to the Integrated Selection (IS) heuristic, by recommending each resource combinations based on the present reservations and resource properties.

## VII. SCALABILITY

It is important to consider the scalability and the scheduling algorithms of the architecture. The architecture may consist of process information (PI), Handling information (HI), Storage Information(SI), and Workflow Engine Service(WE), only if each resource can manage one of each service, as it has been noted that managing multiple services will result into problems. This factor allows each section to be managed by partitioning the manufacturing system, this management is done by dedicated services such as Process Information, Storage Information, Handling Information and Workflow Engine services, the approach thereby helps to spread the load across different resources. A scenario of multiple scheduler in the system is possible due to the high level reservation system of the infrastructure. Howbeit, it is important to be careful to avoid a disagreement as a result of competition between the schedulers when they have same resource reserved by both at the same time. In the architecture, the components presents such as SI, HI and PI, are to ensure the acceptance of one reservation presented by the scheduler and also to give report of the failure or success of the reservation presented.

## VIII. THE CASE STUDY

A randomly created different product workflows is used as a case study, with random numbers of 250, 500 and 750 chosen. This amount will be produce in different five stations in a manufacturing system, with each stations having one buffer, one conveyor and five machines. We assume that machines positioned in each stations can achieve the same manufacturing task. Only the conveyors and buffers at each stations can receive a capacity of four parts. There exist the main conveyor with capacity of ten pallet, which enables material to be handle among the stations. Figure 4 presented the schematic layout of the architecture.

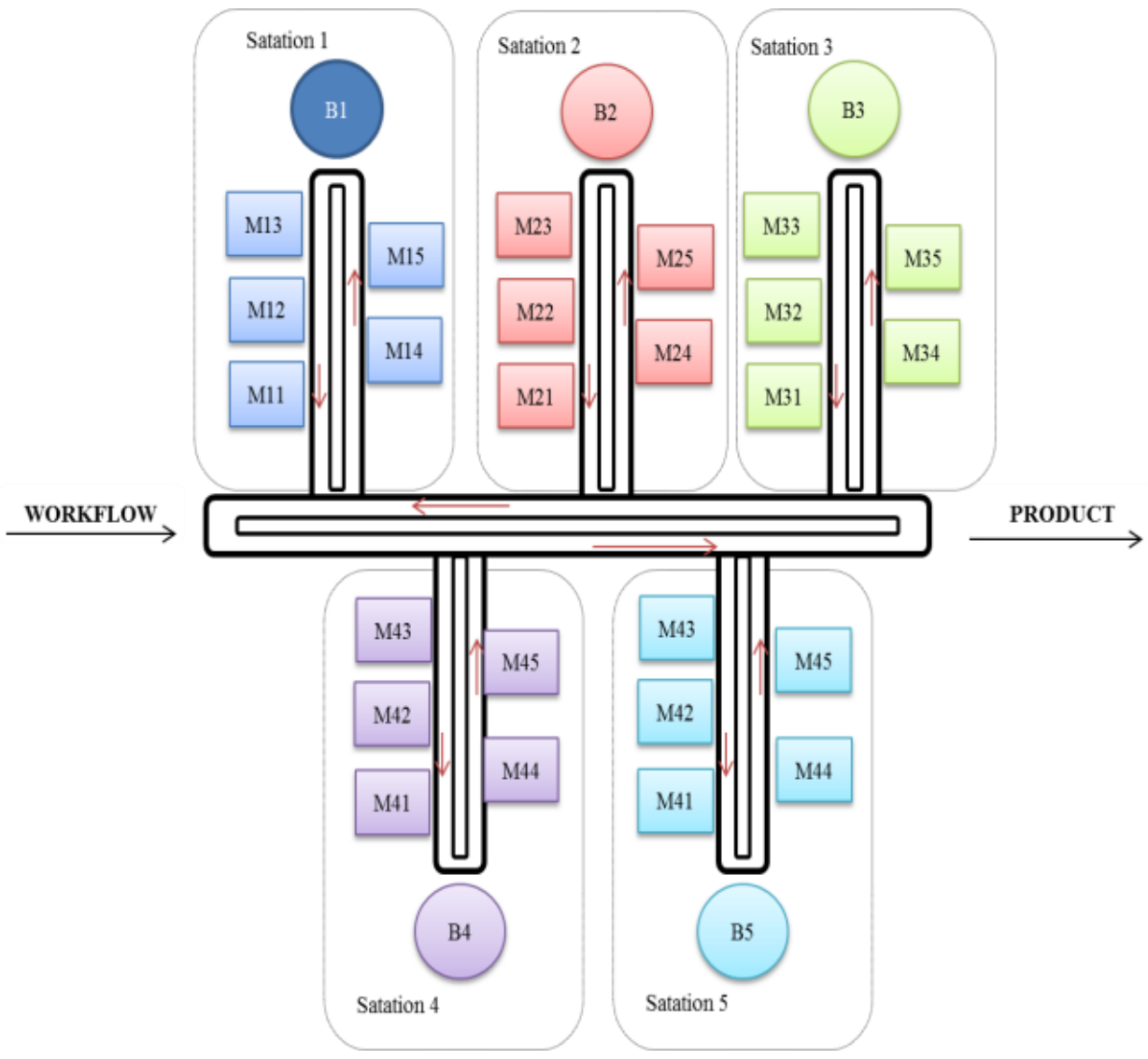

Figure 4 The schematic layout of the system

The scheduler is giving two different order to carry out; the first order is to gather information, such as duration of machining. For example, Station A is needed by product A in 20s, product A is needed in Station B in 50s, station C in 100s,station D in 180s and station E in 200s. Deadline to customer's request and sequence of the machining is also part of the order assigned to the scheduler; this type of order is giving to off-line scheduler. The second order giving to the scheduler is the time of arrival of product requested by the scheduler the online scheduler of the architecture is been activated by the arrival time. Results from twenty simulations done differently was presented in Fig 5 and 5b, this results shows the average CP work flow time for execution and waiting time for scheduling algorithm in offline mode and completed assignment rate of workflow-to-resource. The result of Figure Lu-IS present a more developed rate compared to others. In other word, Fig Lu-IS has a lesser waiting time compared to other algorithm, it also has a better execution time. The worse results in terms of execution and waiting time, CP workflow-to-resource rate of assignment was giving by CD-DS.

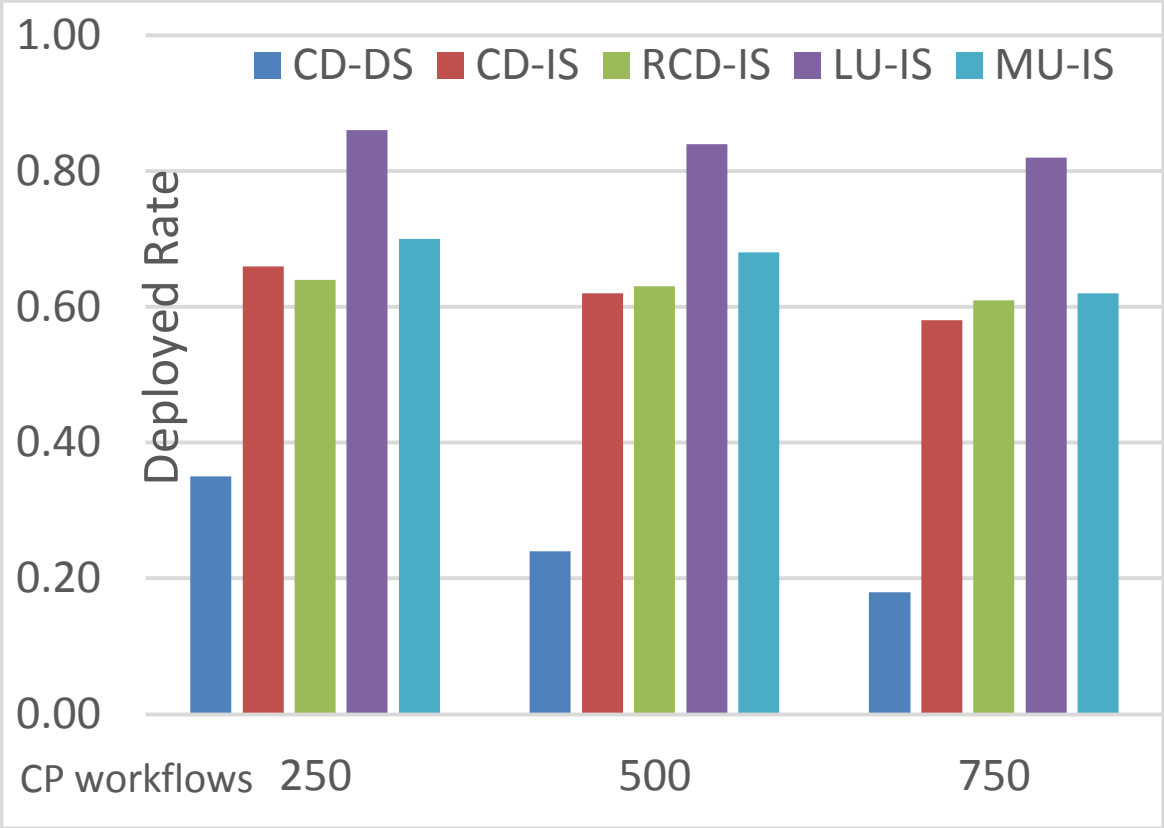

Figure 5a Simulation results

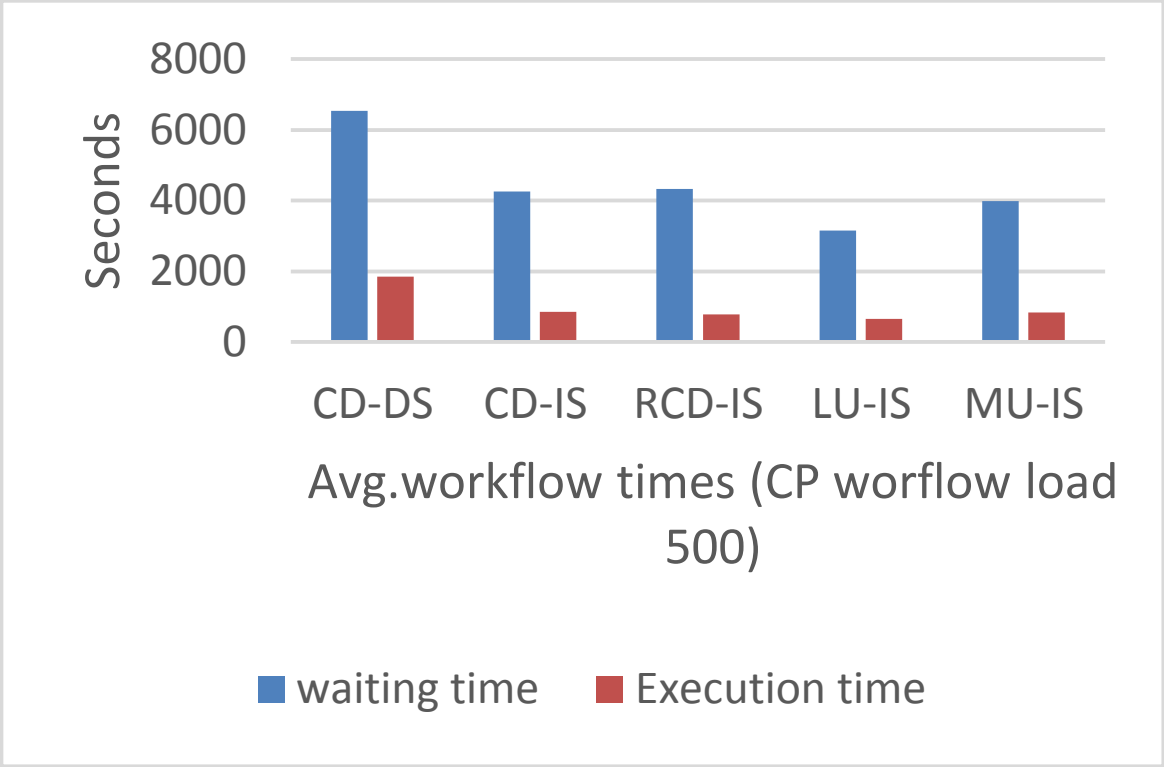

Figure 5a Simulation results

A successful rate of deployment for both IS and DS algorithms in online mode was presented in Fig 6(a),while the average CP workflow execution and waiting time was shown in Fig 6(b). This results were obtained from twenty simulations. From these results, the assignment rate and the average execution and waiting time of DS is poor, very low to the extent of considering it useless.

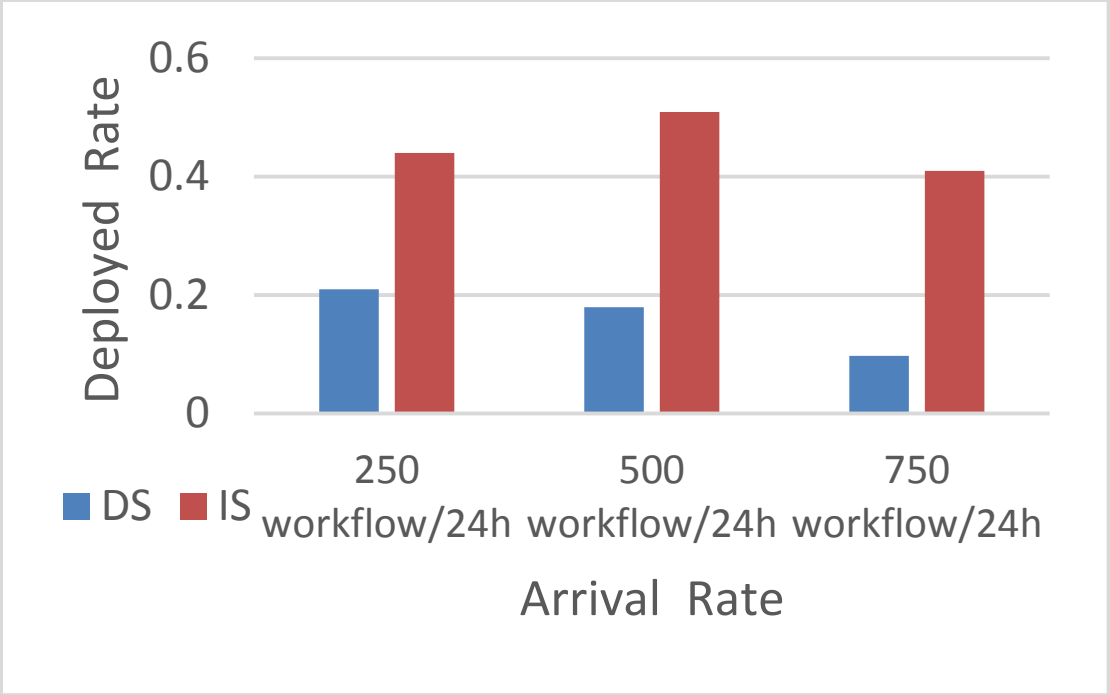

Figure 6a Deployment rates for both DS and IS online algorithms

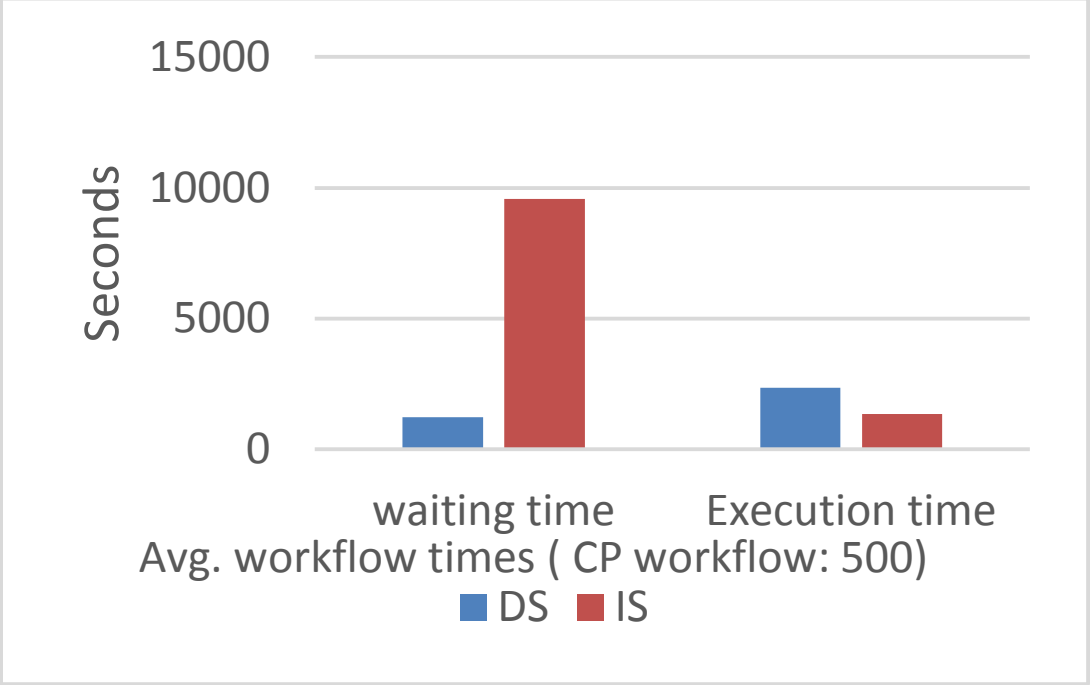

Figure 6b CP workflow waiting time and execution time

## IX. CONCLUSION

In conclusion, one important manufacturing approach is Mass customization. In the current market globalization, quick response to demand and agility has become very important features in most companies, these features also help for intense competition and for rapid implementation of new technologies. Mass customization is the enablement to provide products designed for individuals to users in mass-market economy. A service Oriented Architecture and its peculiarities was presented. The architecture uses high level reservations to improve efficiency by eliminating possible problems as a result of congestion. To improve the use of resources, both offline and online algorithms were combined. The result of implementing such algorithms was presented, which contributed beneficially for lower workflow waiting time and higher flow-to-resource rate of assignment, it also helps the scalability.